\documentclass{sola}

\usepackage{graphicx}
\usepackage{mslapa}
\usepackage{url} %this package should fix any errors with URLs in refs.
\usepackage{amsmath}
\usepackage{braket}

\usepackage{changes}
\usepackage{cancel}
\usepackage{ulem}

\begin{document}

%%%%%%%%%%%%%%%%%%%%%%%%%%%%%%%%%%%%%%%%%%%%%%%%%%%%%%%%%%%%%%%%%%%%%
% TITLE
%%%%%%%%%%%%%%%%%%%%%%%%%%%%%%%%%%%%%%%%%%%%%%%%%%%%%%%%%%%%%%%%%%%%%
%
% May use \\ to break lines in title

\title{Quantum Algorithm for a Stochastic Multicloud Model}

%%%%%%%%%%%%%%%%%%%%%%%%%%%%%%%%%%%%%%%%%%%%%%%%%%%%%%%%%%%%%%%%%%%%%
% AUTHORS
%%%%%%%%%%%%%%%%%%%%%%%%%%%%%%%%%%%%%%%%%%%%%%%%%%%%%%%%%%%%%%%%%%%%%
%

\author{Kazumasa Ueno\affil{1} and Hiroaki Miura\affil{1}}

\affiliation{1}{Department of Earth and Planetary Science, Graduate School of Science, The University of Tokyo, Tokyo, Japan}
% \affiliation{2}{Affiliation A, Tsukuba, Japan}
% \affiliation{3}{Affiliation B, Tokyo, Japan}

%\correspondingauthor{Name}{Institution}{Address}{Email address}
\correspondingauthor{Kazumasa Ueno}{Department of Earth and Planetary Science, Graduate School of Science, The University of Tokyo}{7-3-1 Hongo, Bunkyo-ku, Tokyo, Japan}{kazumasa-e67@eps.s.u-tokyo.ac.jp}

%%%%%%%%%%%%%%%%%%%%%%%%%%%%%%%%%%%%%%%%%%%%%%%%%%%%%%%%%%%%%%%%%%%%%
% RUNNING TITLE
%%%%%%%%%%%%%%%%%%%%%%%%%%%%%%%%%%%%%%%%%%%%%%%%%%%%%%%%%%%%%%%%%%%%%
%
% The running title should encapsulate the article and not exceed 10 words.

\runningtitle{Quantum Algorithm for a Stochastic Multicloud Model}
\runningauthor{Ueno and Miura}

% \linenumbers

%%%%%%%%%%%%%%%%%%%%%%%%%%%%%%%%%%%%%%%%%%%%%%%%%%%%%%%%%%%%%%%%%%%%%
% ABSTRUCT
%%%%%%%%%%%%%%%%%%%%%%%%%%%%%%%%%%%%%%%%%%%%%%%%%%%%%%%%%%%%%%%%%%%%%
%

\begin{abstract}
  Quantum computers have attracted much attention in recent years. This is because the development of the actual quantum machine is accelerating. Research on how to use quantum computers is active in the fields such as quantum chemistry and machine learning, where vast amounts of computation are required. However, in weather and climate simulations, less research has been done despite similar computational demands. In this study, a quantum computing algorithm is applied to a problem of the atmospheric science. The effectiveness of the proposed algorithm is evaluated using a quantum simulator. The results show that it can achieve the same simulations as a conventional algorithm designed for classical computers. More specifically, the stochastically fluctuating behavior of a multi-cloud model was obtained using classical Monte Carlo method, and comparable results are also achieved by utilizing probabilistic outputs of computed quantum states. Our results show that quantum computers have a potential to be useful for the atmospheric and oceanic science, in which stochasticity is widely inherent.
\end{abstract}

%%%%%%%%%%%%%%%%%%%%%%%%%%%%%%%%%%%%%%%%%%%%%%%%%%%%%%%%%%%%%%%%%%%%%
% MAIN TEXT
%%%%%%%%%%%%%%%%%%%%%%%%%%%%%%%%%%%%%%%%%%%%%%%%%%%%%%%%%%%%%%%%%%%%%
%

\section{Introduction}
\label{sec:introduction}

Quantum computers are now gaining interest around the world. This trend can be seen especially in quantum chemistry 
\shortcite
(e.g.,)
{Bauer2020,Cao2019,Motta2022}, optimization 
\shortcite
(e.g.,)
{Grover1996,Hastings2018,Moll2018}, and quantum machine learning \shortcite
(e.g.,)
{Biamonte2017,Mitarai2018}. 
Recent research on the quantum algorithms may be divided into two groups. One is the research on noisy intermediate scale quantum (NISQ) algorithms \shortcite(e.g.,){Preskill2018,Bharti2022}. The NISQ computers are quantum computers that we can access at present, but are threatened by noise since they do not have a tolerance for errors. Despite this lack of error tolerance, they are expected to calculate the problems that we have not solved on classical computers.
The other is algorithms of fault-tolerant quantum computers (FTQCs). The FTQC will be equipped with millions of qubits and can eliminate the effects of errors. These days many types of research are conducted also in the fields of machine development and error correction. Significant improvements have been made recently in the development of the FTQCs \shortcite(e.g.,){Google-Quantum-AI2023,Bluvstein2024}, which indicate it is important to prepare for the era of the FTQCs.

In the field of weather and climate science, however, little research has been done on the usage of quantum computers. One possible reason is that we know less about the problem types that are suitably applicable to quantum computers. 
Tennie and Palmer \citeyear{Tennie2023} have used a quantum algorithm on an emulator to solve nonlinear differential equations. They have argued that encoding and calculating enormous variables in the weather and climate simulations are well-suited for quantum computers. However, due to the readout problems, such usage does not seem to work currently.

Parameterizations used in atmospheric and oceanic models to incorporate influences of unresolved phenomena may be a problem that can circumvent limitations due to the readout problem. Typically, the input and output values in parameterizations are the cell means. The atmospheric radiation, for example, takes much computational cost for the accurate calculation to consider clouds overlap when the Monte Carlo method is used 
\shortcite{Marshak2005}. If an efficient quantum algorithm for the radiation processes is developed, the computational cost will be reduced, and the calculation will be more precise.

Stochasticity is also essential for atmospheric and oceanic phenomena, and is well-suited for an efficient use of quantum computers as Tennie and Palmer \citeyear{Tennie2023} noted. Berner et al. \citeyear{Berner2017} argued that stochastic parameterizations are essential for improving the predictability of weather and climate models because they can represent variabilities of subgrid-scale states, such as convection and gravity waves. They argued that employing stochastic and statistical approaches can partly represent the uncertainty of climate and weather systems. As a method to include complementing variabilities, a superparameterization was suggested by Grabowski and Smolarkiewicz \citeyear{Grabowski1999}, but this requires further more computational time than usual parameterizations. Novel quantum algorithm to represent the probability changes of subgrid-scale can be a clue to resolve uncertainty in weather and climate predictions through examining vast amount of stochasticity and statistics.

In this study, we report an example of problems that quantum computers can be used in the atmospheric science. The stochastic behavior of clouds was chosen as the target. We used the stochasticity of quantum computers to represent the stochastic aspects of clouds. More specifically, a stochastic multi-cloud model \shortcite{Khouider2010} was chosen for the application.
The stochastic multi-cloud model is a possible approach for representing stochasticity that cannot be captured by the dynamical cores of General Circulation Models (GCMs). The utility of the stochastic multi-cloud model has been further investigated, particularly in the tropical atmosphere \shortcite{Dorrestijn2016,Goswami2017,Khouider2023,Karsten2017}. Although the stochastic multi-cloud model is simple, it encompasses important aspects, making it a suitable choice for implementing a quantum algorithm as a first step.

\section{Stochastic Multicloud Model and Quantum Algorithm}

\subsection{Stochastic Multicloud Model}

A brief summary of the Stochastic Multicloud Model (SMCM) proposed by 
Khouider et al. \citeyear{Khouider2010} 
is as follows. In the SMCM, the computing domain is subdivided into a lattice. Developments and decays of clouds are assigned to each lattice site. The lattice scale is usually $O (1)$-km. Each lattice site is independent of the others and only depends on the environment. 
% The original domain represents the state of large-scale dynamics, which is reffered to as environment.
The environment refers to the state of the large-scale dynamics, which is represented by the original domain. 
Although interactions between lattices are considered in Khouider \citeyear{Khouider2014}, we have omitted them here because our aim is not to investigate cloud interactions, but to utilize quantum computers for cloud representations. 
In addition to the state of clear sky, three types of clouds are represented within each lattice site: congestus cloud, deep convective cloud, and stratiform anvil. The states of each lattice are indexed as
\begin{equation}
      X_t^i = 
      \begin{cases}
            0 &\text{if the $i_{th}$ site is clear sky,} \\
            1 &\text{if the $i_{th}$ site is occupied by the congestus cloud,} \\
            2 &\text{if the $i_{th}$ site is occupied by the deep convective cloud,} \\
            3 &\text{if the $i_{th}$ site is occupied by the stratiform anvil,}
      \end{cases}
\end{equation}
where $i=1,\ldots,N$ is the $i_{th}$ lattice site, $N$ is the number of lattice sites in the original grid, and $t$ is time.

The time evolution of each site is assumed to obey a Markov process, which is a continuous-time stochastic process. In the Markov process, only the current lattice state affects the state one step later. The state change is represented by transition probabilities, which are defined as follows:
\begin{equation}
      \begin{split}
      P_{lk}^i &= Prob\{X_{t+\Delta t}^i=k | X_t^i=l\} \\ 
      &= 
      \begin{cases}
            R_{lk}^i\Delta t + o(\Delta t) & \text{for}\ l,k=0,1,2,3, \text{and}\ l\neq k, \\
            1-\sum_{k=0,k\neq l}^{3} P_{lk}^i   & \text{for}\ l,k=0,1,2,3, \text{and}\ l=k,
      \end{cases}
      \end{split}
\end{equation}
where $l$ and $k$ are states of $i_{th}$ lattice site on time levels $t=t$ and $t=t+\Delta t$ respectively, $\Delta t>0$ is a small time increment, the $R_{lk}^i$ are transition rates, and $o(\Delta t)$ means higher order infinitesimals of $\Delta t$.
Here, $R_{lk}^i$ is assumed to depend on the Convective Available Potential Energy (CAPE) and dryness of the environment. Some intuitive interaction rules are applied to determine the form of $R_{lk}^i$. For example, it is reasonable to assume that clear sky sites are unlikely to become stratiform anvil sites within a short time increment. See 
Khouider et al. \citeyear{Khouider2010} 
for details on the rules. The final forms of $R_{lk}^i$ are expressed as follows: 
\begin{align*}
      R_{01}&=\frac{1}{\tau_{01}}\Gamma(C)\Gamma(D), &
      R_{02}&=\frac{1}{\tau_{02}}\Gamma(C)(1-\Gamma(D)), \\
      R_{10}&=\frac{1}{\tau_{10}}\Gamma(D), &
      R_{12}&=\frac{1}{\tau_{12}}\Gamma(C)(1-\Gamma(D)), \\
      R_{20}&=\frac{1}{\tau_{20}}(1-\Gamma(C)), &
      R_{23}&=\frac{1}{\tau_{23}},\qquad
      R_{30}=\frac{1}{\tau_{30}}, \\
      R_{03}&=R_{13}=R_{21}=R_{31}=R_{32}=0,
\end{align*}
where $\tau_{lk}$ is the time scale of formation or decay of each cloud type or conversion of one cloud type to another, and $C$ and $D$ are CAPE and dryness ratio, respectively. $\Gamma$ is a function expressed by 
\begin{equation}
      \Gamma(x)=
      \begin{cases}
            1-e^{-x} &\text{if $x>0$,} \\
            0 &\text{otherwise.}
      \end{cases}
\end{equation}
The values of $\tau_{lk}$ are listed in Table \ref{table:tau}. The fractions of each state in the original domain are denoted as $\sigma_{cs}$ for clear sky, $\sigma_{c}$ for congestus cloud, $\sigma_{d}$ for deep convective cloud, and $\sigma_{s}$ for stratiform anvil, respectively. We compute the time variation of these state fractions.

Two reference solutions are created by conventional algorithms to verify the correctness of the solution obtained by our new quantum algorithm. One is by a deterministic calculation as follows: 
\begin{equation}
      \label{sigma_calc}
      \begin{pmatrix}
            \sigma_{cs}^{t+\Delta t} \\
            \sigma_{c}^{t+\Delta t} \\
            \sigma_{d}^{t+\Delta t} \\
            \sigma_{s}^{t+\Delta t} \\
      \end{pmatrix}
      =
      \begin{pmatrix}
            P_{00} & P_{10} & P_{20} & P_{30} \\
            P_{01} & P_{11} & P_{21} & P_{31} \\
            P_{02} & P_{12} & P_{22} & P_{32} \\
            P_{03} & P_{13} & P_{23} & P_{33} \\
      \end{pmatrix}
      \begin{pmatrix}
            \sigma_{cs}^{t} \\
            \sigma_{c}^{t} \\
            \sigma_{d}^{t} \\
            \sigma_{s}^{t} \\
      \end{pmatrix}, \qquad P_{lk}=\frac{1}{N}\sum_{i=1}^{N}P_{lk}^i.
\end{equation}
The other is by a Monte Carlo calculation. The acceptance-rejection algorithm is used for the Monte Carlo calculations in the same way as 
Khouider et al. \citeyear{Khouider2010}
.

\subsection{Quantum Algorithm}
Quantum computers can perform matrix calculations. By utilizing this ability to calculate algebraic operations in (\ref{sigma_calc}) and leveraging the characteristics of stochastical outputs, we can calculate the SMCM. Note that the basics of quantum information are summarized in the supplementary material.
The details of our quantum algorithm for calculating the SMCM are as follows. The calculations are performed in the form of (\ref{sigma_calc}), but some constraints are needed on quantum computing systems. In quantum computing systems, vectors must be normalized, and operators must be in the form of unitary matrices. In (\ref{sigma_calc}), the normalization of the state fraction vector is straightforward. However, the operator matrix is not always unitary. To solve this problem, we used the method of the linear combination of unitaries (LCUs), as described by 
Xin et al. \citeyear{Xin2020}. Denote the matrix in the right-hand side of (\ref{sigma_calc}) as $A$. $A$ is a normalized matrix satisfying $\|A\|\le 1$, where $\|*\|$ represents the spectral norm, corresponding to the largest singular value of the matrix $*$. 
Thus, $A$ can be decomposed into the symmetric and asymmetric matrices as 
\begin{equation}
      \label{eq:BCandA}
      B=\frac{1}{2}\left(A + A^{\dag}\right), \qquad C=\frac{1}{2i}\left(A-A^{\dag}\right),
\end{equation}
where $A^{\dag}$ is the adjoint matrix of $A$. By introducing unitary matrices $F_1, F_2, F_3,$ and $F_4$ as
\begin{equation}
      \begin{split}
      F_1 &= B+i\sqrt{I-B^2},\qquad F_2 = B-i\sqrt{I-B^2}, \\
      F_3 &= iC-\sqrt{I-C^2},\qquad F_4 = iC+\sqrt{I-C^2},
      \end{split}
\end{equation}
$B$ and $C$ can be rewritten as
\begin{equation}
      \label{eq:BCandFs}
      B = \frac{1}{2}\left(F_1+F_2\right), \quad \text{and} \quad C=\frac{1}{2}\left(F_3+F_4\right).
\end{equation}
From (\ref{eq:BCandA}) and (\ref{eq:BCandFs}), $A$ can be expressed in a summation of unitary matrices as
\begin{equation}
      \label{eq:AandFs}
      A = \frac{1}{2}\left(F_1+F_2+F_3+F_4\right).
\end{equation}

% \begin{center}
%       \Qcircuit @C=1em @R=.7em {
%         &\lstick{\ket{0}_{a_0}} & \gate{H} & \ctrlo{1} & \ctrlo{1} & \ctrl{1} & \ctrl{1} & \gate{H} &  \meter \\
%         &\lstick{\ket{0}_{a_1}} & \gate{H} & \ctrlo{1} & \ctrl{1} & \ctrlo{1} & \ctrl{1} & \gate{H} &  \meter \\
%         &\lstick{\ket{0}_{q_0}} & \multigate{1}{U_{init}}  & \multigate{1}{F_1} & \multigate{1}{F_2} & \multigate{1}{F_3} & \multigate{1}{F_4} & \qw & \meter \\
%         &\lstick{\ket{0}_{q_1}} & \ghost{U_{init}}      & \ghost{F_1} & \ghost{F_2} &  \ghost{F_3} & \ghost{F_4}   & \qw & \meter 
%       }
% \end{center}

\begin{figure}[htbp]
  \begin{center}
        % pi_1とpi_2とpi_3の線を入れたい
      %   \includegraphics[width=0.8\linewidth,natwidth=80.76,natheight=27.72]{./fig/circuit-1200.png}
        \includegraphics[width=0.8\linewidth]{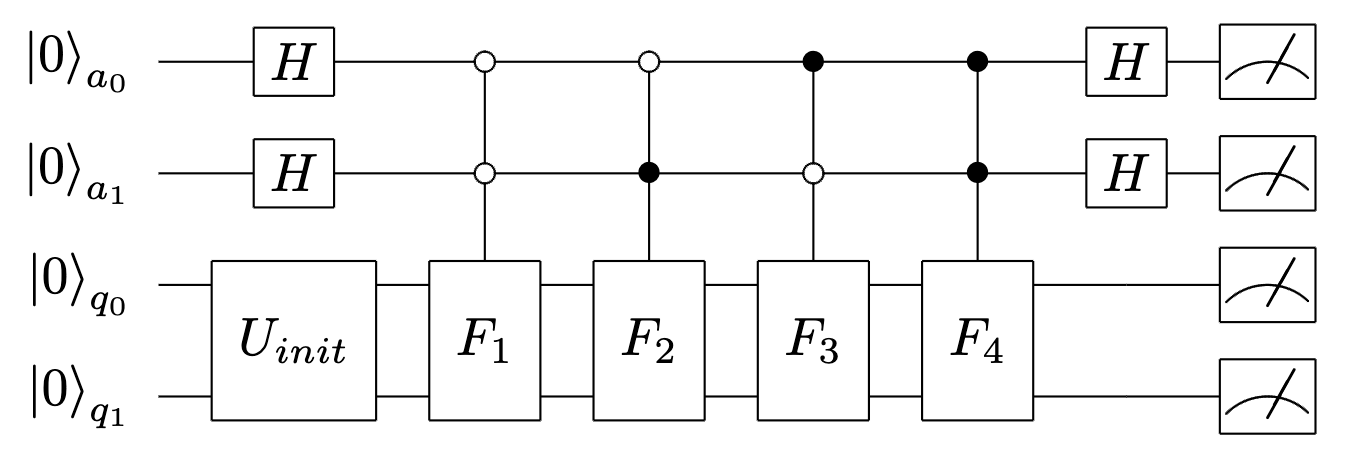}
        \caption{Quantum circuit for the Stochastic Multicloud Model (SMCM). Horizontal lines express qubits, and boxes are operations of unitary matrices for these qubits. $H$ is the hadamard gate, $U_{init}$ is an initializing unitary gate, and $F_1, F_2, F_3,$ and $F_4$ are controlled unitary gates. The right boxes with meters mean observation. In order to obtain the cloud fraction values, calculations of this circuit must be conducted again and again.}
        \label{qcircuit}
  \end{center}
\end{figure}

Figure \ref{qcircuit} is the quantum circuit we made for the SMCM. The top two qubits are ancilla qubits ($a_0,a_1$), which are extra bits for the calculation. These bits are introduced to support the summation of matrices in (\ref{eq:AandFs}), which is realized by implementing Hadamard gates. The main calculation is conducted on the bottom two qubits ($q_0,q_1$). All qubits are initialized as $\ket{0}$. First, we create superposition states by operating the Hadamard gates on the ancilla qubits. At the same stage, we initialize the main qubits by operating $4\times 4$ unitary matrix that holds the normalized cloud fraction vector on the first column, that is,
\begin{equation}
      U_{init} = \frac{1}{\sqrt{(\sigma_{cs}^t)^2+(\sigma_{c}^t)^2+(\sigma_{d}^t)^2+(\sigma_{s}^t)^2}}\begin{pmatrix}
            \sigma_{cs}^t & * & * & * \\
            \sigma_{c}^t & * & * & * \\
            \sigma_{d}^t & * & * & * \\
            \sigma_{s}^t & * & * & *
      \end{pmatrix},
\end{equation}
where $*$'s are arbitrary elements that make $U_{init}$ unitary.
The qubit states denoted by $\ket{\pi_1}$ after this operation can be written as
\begin{equation}
      \ket{\pi_1} = \frac{1}{2}\left(\ket{00}_{a_0a_1}+\ket{01}_{a_0a_1}+\ket{10}_{a_0a_1}+\ket{11}_{a_0a_1}\right)\otimes \ket{\sigma^t}_{q_0q_1},
\end{equation}
where $\otimes$ is the Kronecker product, and $\ket{\sigma^t}$ is the cloud fraction ket vector with four components, that is,
\begin{equation}
      \label{eq:sigma_vector}
      \ket{\sigma^t} = \frac{1}{\sqrt{(\sigma_{cs}^t)^2+(\sigma_{c}^t)^2+(\sigma_{d}^t)^2+(\sigma_{s}^t)^2}}\begin{pmatrix}
            \sigma_{cs}^t \\ \sigma_{c}^t \\ \sigma_{d}^t \\ \sigma_{s}^t
      \end{pmatrix}.
\end{equation}
Next, controlled unitary gates are operated on each state as
\begin{multline}
      \ket{\pi_2} = \frac{1}{2}\left(\ket{00}_{a_0a_1}\otimes F_1\ket{\sigma^t}_{q_0q_1}+\ket{01}_{a_0a_1}\otimes F_2\ket{\sigma^t}_{q_0q_1} \right.\\
      \left. +\ket{10}_{a_0a_1}\otimes F_3\ket{\sigma^t}_{q_0q_1}+\ket{11}_{a_0a_1}\otimes F_4\ket{\sigma^t}_{q_0q_1}\right).
\end{multline}
Finally, the Hadamard gates are operated on the ancilla qubits again to obtain the next step $\sigma$ which is encoded in the state where the ancilla qubits are $\ket{00}$. This state can be written as
\begin{equation}
      \label{eq:pi3}
      \ket{\pi_3} = \frac{1}{2} \ket{00}_{a_0a_1} \otimes \underbrace{\frac{1}{2}\left(F_1+F_2+F_3+F_4\right) \ket{\sigma^t}_{q_0q_1}}_{\ket{\sigma^{t+\Delta t}}} + \ket{\lambda}_{a_0a_1q_0q_1},
\end{equation}
where $\ket{\lambda}$ is a state that we are not interested in.
When we observe the quantum states calculated as $\ket{\pi_3}$, we obtain 4-bit strings representing classical states. The probability of observing each state is proportional to the square of the calculated quantum amplitude. For example, the probability of observing the $\ket{0000}$ state is 
\begin{align}
      \nonumber
      p_{0000} &= |\braket{0000|\pi_3}_{a_0a_1q_0q_1}|^2 \\
      \nonumber
      &= \left|\frac{1}{2}\braket{00|00}_{a_0a_1}\otimes \braket{00|\sigma^{t+\Delta t}}_{q_0q_1}\right|^2 \\
      \label{eq:p0000}
      &= \frac{1}{4}\frac{(\sigma_{cs}^{t+\Delta t})^2}{(\sigma_{cs}^{t+\Delta t})^2+(\sigma_{c}^{t+\Delta t})^2+(\sigma_{d}^{t+\Delta t})^2+(\sigma_{s}^{t+\Delta t})^2}.
\end{align}
In the calculation of (\ref{eq:p0000}), we used the relation of (S4), (S7), and (S8) in the supplementary, such as
\begin{align}
      \braket{00|00} &= \begin{pmatrix}
            1 & 0 & 0 & 0
      \end{pmatrix}\begin{pmatrix}
            1 \\ 0 \\ 0 \\ 0
      \end{pmatrix} = 1\\
      \nonumber
      \braket{00|\sigma^{t+\Delta t}} &= \frac{1}{\sqrt{(\sigma_{cs}^{t+\Delta t})^2+(\sigma_{c}^{t+\Delta t})^2+(\sigma_{d}^{t+\Delta t})^2+(\sigma_{s}^{t+\Delta t})^2}} \begin{pmatrix}
            1 & 0 & 0 & 0
      \end{pmatrix}
      \begin{pmatrix}
            \sigma_{cs}^{t+\Delta t} \\ \sigma_{c}^{t+\Delta t} \\ \sigma_{d}^{t+\Delta t} \\ \sigma_{s}^{t+\Delta t}
      \end{pmatrix} \\
      &= \frac{\sigma_{cs}^{t+\Delta t}}{\sqrt{(\sigma_{cs}^{t+\Delta t})^2+(\sigma_{c}^{t+\Delta t})^2+(\sigma_{d}^{t+\Delta t})^2+(\sigma_{s}^{t+\Delta t})^2}}.
\end{align}
In a similar manner, since $\ket{01}=\begin{pmatrix}0 & 1 & 0 & 0\end{pmatrix}^T$ and
\begin{align}
      \braket{01|\sigma^{t+\Delta t}} = \frac{\sigma_{c}^{t+\Delta t}}{\sqrt{(\sigma_{cs}^{t+\Delta t})^2+(\sigma_{c}^{t+\Delta t})^2+(\sigma_{d}^{t+\Delta t})^2+(\sigma_{s}^{t+\Delta t})^2}},
\end{align}
the probability of observing the $\ket{0001}$ state is
\begin{align}
      p_{0001} = \frac{1}{4}\frac{(\sigma_{c}^{t+\Delta t})^2}{(\sigma_{cs}^{t+\Delta t})^2+(\sigma_{c}^{t+\Delta t})^2+(\sigma_{d}^{t+\Delta t})^2+(\sigma_{s}^{t+\Delta t})^2}.
\end{align}
The probabilities of observing the $\ket{0010}$ and $\ket{0011}$ states are proportional to $(\sigma_d^{t+\Delta t})^2$ and $(\sigma_s^{t+\Delta t})^2$, respectively. If the first two qubits are not $\ket{00}$, the probabilities of observing those states are out of our interests since they corresponds to the garbage states.

We derive a statistical distribution through sampling, which allows us to estimate the squares of the cloud fraction states. By taking their square roots and normalizing them to make the sum equals 1, cloud fraction states at time $t+\Delta t$ are obtained. This process is conducted on a classical computer after the sampling at time $t$ has been completed. Note that repeated calculations (observations) are necessary at each time step in this approach. The number of repetitions is called the shot number here. 

We used qiskit Aer \cite{Qiskit} simulator to compute the above algorithm.
% Though we wanted to calculate on real quantum devices, we did not because, as of December 2023, controlled-Unitary operators were unavailable for the specification of primitives, a framework allowing iterative quantum calculations.
Simulation settings are similar to those described by Khouider et al. \citeyear{Khouider2010}. The initial fractions of $\sigma_{cs}, \sigma_c, \sigma_d,$ and $\sigma_s$ were set equally at 0.25. The values of CAPE, $C=0.25$, and dryness, $D=0.75$, are used throughout the entire duration of the simulation. The integration is performed over 100 hours. Until this time, the cloud fractions reached their quasi-equilibrium states (Figure \ref{frac_change}).

\section{Results}
\label{sec:results}

\begin{figure}[htbp]
  \begin{center}
        \includegraphics[width=\linewidth]{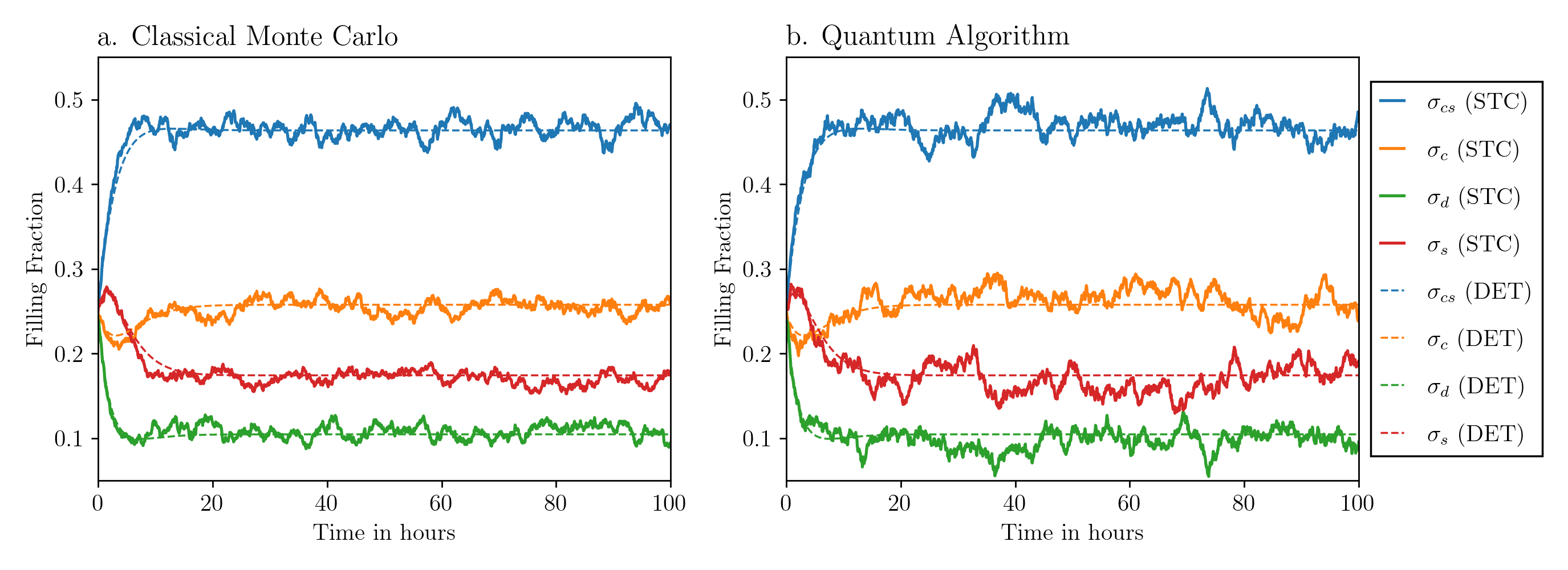}
        \caption{Time evolution of the fractions of clear sky and cloud types. In both panels, the dashed lines represent the results of the same deterministic simulation. The solid lines illustrate the results of the stochastic simulations for (a) the classical Monte Carlo algorithm (with 400 lattice sites), and for (b) the quantum algorithm (with 40,000 shots).}
        \label{frac_change}
  \end{center}
\end{figure}

Figure \ref{frac_change} displays the time evolutions of the fractions of states observed in the SMCM simulations. In the deterministic simulation, the fractions reach their equilibrium values within the first 10 hours. The equilibrium values are as follows: $\sigma_{cs}=0.46, \sigma_c=0.26, \sigma_d=0.10,$ and $\sigma_s=0.17$, respectively. In the case of using the classical Monte Carlo algorithm (Figure \ref{frac_change}a), the results fluctuate around the deterministic results. These results are qualitatively similar to the results of Khouider et al. \citeyear{Khouider2010}. In the case of using the quantum algorithm (Figure \ref{frac_change}b), the results are similar to that obtained with the classical Monte Carlo simulation. The transition behaviors observed in both simulations using the classical Monte Carlo and quantum algorithms closely follow the result of the deterministic simulation.

\begin{figure}[htbp]
  \begin{center}
        \includegraphics[width=\linewidth]{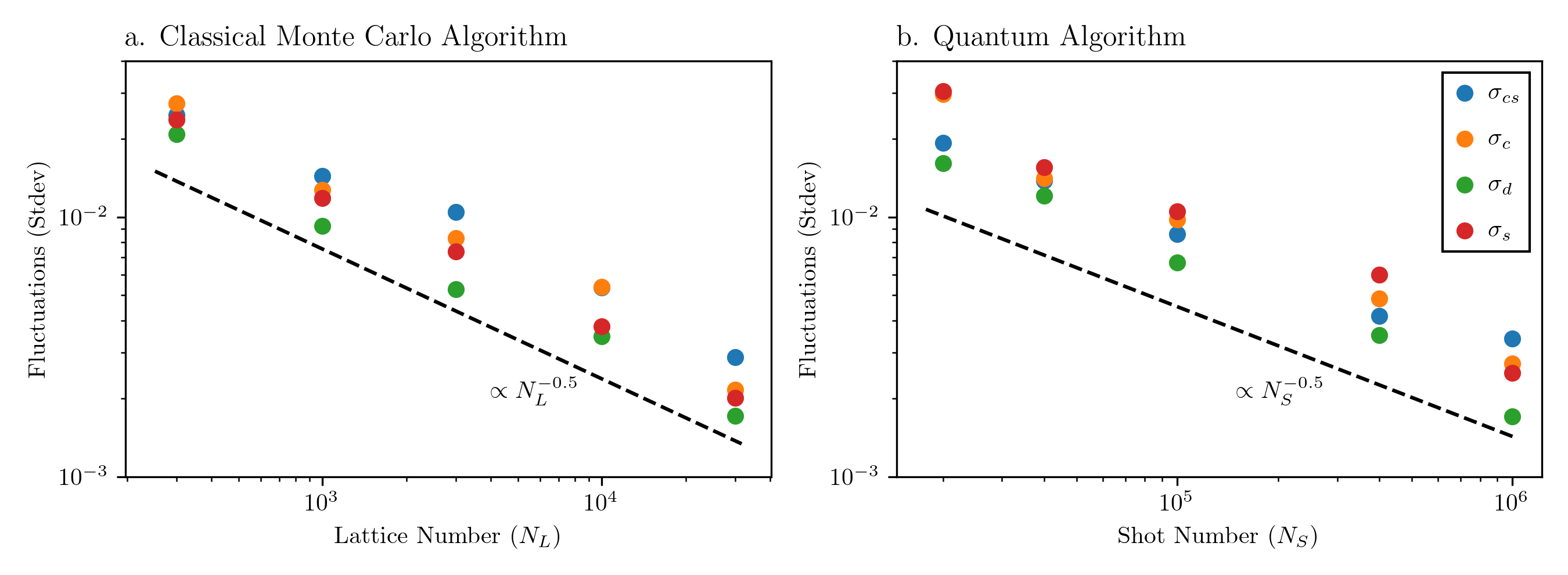}
        \caption{The degrees of fluctuations around the equilibrium states obtained by the deterministic simulation. The root-mean-squares of the stochastic values minus the deterministic value are plotted against (a) the number of lattice sites in the classical Monte Carlo algorithm, and against (b) the shot numbers in the quantum algorithm. The dashed black lines are proportional to $N_L^{-0.5}$ and $N_S^{-0.5}$, where $N_L$ represents the lattice number for the classical Monte Carlo, and $N_S$ represents the number of shots for the quantum algorithm, respectively.}
        \label{fluctuation}
  \end{center}
\end{figure}

To determine the parameter that influences the fluctuation magnitudes, we conducted the same experiment with different parameter settings. Figure \ref{fluctuation} illustrates the dependency of fluctuation magnitudes on (a) the number of lattice sites in the classical Monte Carlo simulation and (b) the shot number in the quantum algorithm. The larger the number of lattice sites in the classical Monte Carlo simulation, or the larger the number of shots in the quantum simulation, the fewer the fluctuations observed. These results align with the line of $N_L^{-0.5}$ and $N_S^{-0.5}$, where $N_L$ represents the lattice number for the classical Monte Carlo, and $N_S$ represents the number of shots for the quantum algorithm, respectively. It means the results are consistent with the central-limit theorem. Approximately $10^5$ shots are required in the simulation using the quantum algorithm to achieve results similar to those obtained from the classical Monte Carlo calculation, which uses $3\times 10^3$ lattice sites. 

Why is there a difference in the number of samples required between the quantum simulation and the classical Monte Carlo simulation? 
The reason for the fact that the quantum algorithm needs larger sample size to obtain similar fluctuations is explained for three resons. 
The first is an influence of garbage states. The simulation using the quantum algorithm samples from $2^4$ possible states. In addition, if we rewrite the final quantum states shown in (\ref{eq:pi3}) as
\begin{equation}
      \ket{\pi_3} = \sqrt{p_0}\ket{00}_{a_0a_1}\otimes \ket{\sigma^{t+\Delta t}}_{q_0q_1} + \sqrt{1-p_0}\ket{\lambda'}_{a_0a_1q_0q_1},
\end{equation}
where $p_0$ is the probability of obtaining the states of interest, and $\ket{\lambda'}$ is the garbage states, which corresponds to $\ket{\lambda}$ in (\ref{eq:pi3}) but normalized so that the square of the coefficient corresponds to the probability.
Then, the actual number of samples for the 4 states is approximately $p_0N_S$. In our experiment, $p_0$ is 0.25, which contributes fourfold to the gap in the number of samples between the classical Monte Carlo simulation and the quantum simulation. We can increase $p_0$ if we use an amplitude amplification \shortcite{Brassard2002}, but it requires greater gate complexity, and thus, is not adopted.
The second is an influence of the encoding way. We encode the probabilistic states by renormalizing the vector (see (\ref{eq:sigma_vector})) and obtain results in squared form. This contributes twice to the gap. 
The last that explains the remaining nearly fourfold gap may be an influence of the characteristic of the used quantum simulator. Here we denote the fluctuation of the Monte Carlo simulation as $C_{MC}N_L^{-0.5}$ and one of the quantum algorithm as $C_{Q}N_S^{-0.5}$. Then, for the Aer simulator, we observe $C_Q/C_{MC}=4$. In our experience, this ratio varies depending on the type of simulator or the actual hardware used.

\section{Discussion and Conclusions}
\label{sec:conclusion}

We found an atmospheric science problem to which a quantum algorithm was applicable, and actually configured a quantum circuit to compute it successfully.
We implemented the algorithm on quantum circuits using Qiskit \cite{Qiskit}. The Qiskit transpiler decomposed these circuits into native circuits with a depth of $O(1000)$. The gate depth means the number of steps to implement all basic gate operations in a quantum circuit. We can roughly estimate the computation cost by multiplying this gate depth by the number of shots. It is important to note that the size of fluctuations in our algorithm is not dependent on the gate depth but is exponentially dependent on the number of shots. In other words, we do not modify the quantum circuit itself, but only adjust the number of shots to change the fluctuations. This means we cannot fully benefit from the quantum characteristic of superposition states when it comes to the adjustment of fluctuation size. In order to reduce the shot number dependency to a level lower than exponential, an amplitude estimation algorithm \shortcite{Brassard2002} might work. This is a future work. Additionally, we did not include the variation of CAPE and dryness because our primary goal here is to explore the application of a quantum computer to an atmospheric problem. Addressing the variation of CAPE and dryness requires developing an effective quantum algorithm to calculate CAPE and dryness from quantum states. This is also a focus of future work.

Although the quantum computer cannot take full advantage of its potential in our current usage, the results of this study suggest two promising possibilities. The first is matrix calculations of transition probabilities. Quantum computation is superior to classical computation for large matrix calculations. For example, the systems of linear differential equations can be solved using quantum linear system algorithms (QLSAs) \shortcite{Berry2017-vv}. One of the main components of QLSAs is calculating matrix inversion, as represented by the HHL algorithm \shortcite{Harrow2009}. QLSAs perform well, particularly when matrices are sufficiently sparse. Probability distributions typically involve multiple dimensions, making their time development challenging to compute, but they are often sparse. Although this study tested only four states and did not use QLSAs, expanding to a larger number of states and using QLSAs could potentially reduce computational costs compared to classical computers. We expect that we could find suitable applications in the cloud droplet formation processes in weather and climate computations.
The second possibility involves utilizing probabilistic outputs from computed quantum states. It may be feasible to model atmospheric phenomena where there is only one realization state that is determined according to a probability distribution.

Additionally, calculations of radiation, cloud microphysics, and turbulence are potential targets for quantum computing. They demand enormous computations and research is starting to be conducted on these topics \shortcite{Igarashi2024,Mukta2023}. In our context, considering local interactions at each lattice site may be beneficial \shortcite{Khouider2014}. Local interactions are relatively easy to handle on quantum computers, making this a promising target for future work.

Quantum computings have a potential to change the standard for computations. We hope that this research will promote further studies on the usage of quantum computers in weather and climate science.

%%%%%%%%%%%%%%%%%%%%%%%%%%%%%%%%%%%%%%%%%%%%%%%%%%%%%%%%%%%%%%%%%%%%%
% ACKNOWLEDGMENTS
%%%%%%%%%%%%%%%%%%%%%%%%%%%%%%%%%%%%%%%%%%%%%%%%%%%%%%%%%%%%%%%%%%%%%
%

\section*{Acknowledgments}

This work was supported by JSPS KAKENHI Grant Numbers 20H0419b, 20H05727, 20H05729, 20H05731, and 23H01243; and International Graduate Program for Excellence in Earth-Space Science (IGPEES), a World-leading Innovative Graduate Study (WINGS) Program, the University of Tokyo.

%%%%%%%%%%%%%%%%%%%%%%%%%%%%%%%%%%%%%%%%%%%%%%%%%%%%%%%%%%%%%%%%%%%%%
% Supplements
%%%%%%%%%%%%%%%%%%%%%%%%%%%%%%%%%%%%%%%%%%%%%%%%%%%%%%%%%%%%%%%%%%%%%
%
% Give capsule summary of each supplementary material.
% If you do not have any supplementary materials, comment out it.

\begin{supplements}

\item The basics of quantum information are summarized.
% \item Additional information of the Data B.
% \item The source code to calculate the statistics given in Section \ref{sec:results}.

\end{supplements}

%%%%%%%%%%%%%%%%%%%%%%%%%%%%%%%%%%%%%%%%%%%%%%%%%%%%%%%%%%%%%%%%%%%%%
% REFERENCES
%%%%%%%%%%%%%%%%%%%%%%%%%%%%%%%%%%%%%%%%%%%%%%%%%%%%%%%%%%%%%%%%%%%%%
%
% Make your BibTex bibliography by using bibtex command

\bibliographystyle{jmsj}
\bibliography{bibtex/ref}

%%%%%%%%%%%%%%%%%%%%%%%%%%%%%%%%%%%%%%%%%%%%%%%%%%%%%%%%%%%%%%%%%%%%%
% FIGURES
%%%%%%%%%%%%%%%%%%%%%%%%%%%%%%%%%%%%%%%%%%%%%%%%%%%%%%%%%%%%%%%%%%%%%
%

%%%%%%%%%%%%%%%%%%%%%%%%%%%%%%%%%%%%%%%%%%%%%%%%%%%%%%%%%%%%%%%%%%%%%
% TABLES
%%%%%%%%%%%%%%%%%%%%%%%%%%%%%%%%%%%%%%%%%%%%%%%%%%%%%%%%%%%%%%%%%%%%%
%
\begin{table}[htbp]
  \centering
  \begin{tabular}{ccc}
        & description & value \\ \hline
        $\tau_{01}$ & decay of congestus & 1 hour \\
        $\tau_{10}$ & decay of congestus & 5 hours \\
        $\tau_{12}$ & conversion of congestus to deep & 1 hour \\
        $\tau_{02}$ & formation of deep & 2 hours \\
        $\tau_{23}$ & conversion of deep to stratiform & 3 hours \\
        $\tau_{20}$ & decay of deep & 5 hours \\
        $\tau_{30}$ & decay of stratiform & 5 hours \\ \hline
  \end{tabular}
  \caption{The values of $\tau_{lk}$, used in this calculation. These values correspond to case 1 of 
  Khouider et al. (2010)
  .}
  \label{table:tau}
\end{table}

\end{document}